\documentclass[preprint,aps,rsi,showpacs,superscriptaddress]{revtex4-1}
\usepackage{graphicx,epsfig}
\usepackage{epstopdf}
\usepackage{bm}

\usepackage{caption}
\usepackage{subcaption}

\makeatletter

\renewcommand*{\p@subsection}{}

\renewcommand*{\p@subsubsection}{}
\makeatother

\newcommand{\3}{$^3$He}
\newcommand{\4}{$^4$He}

\newcommand{\um}{$\mu$m}
\begin{document}

\title{Apparatus to visualize flows in superfluid $^4$He below 1\,K}

\author{I. Skachko}
\thanks{Present address: UKRI-STFC, Daresbury Laboratory, Warrington WA4 4AD, UK}
\affiliation{Department of Physics and Astronomy, The University of Manchester, Manchester M13 9PL, UK}
\author{J. A. Hay}
\thanks{Present address: Maxeler Technologies, London W6 0ND, UK}
\affiliation{Department of Physics and Astronomy, The University of Manchester, Manchester M13 9PL, UK}
\author{C. O. Goodwin}
\thanks{Present address: School of Mathematical Sciences, University of Nottingham, Nottingham NG7 2RD, UK}
\affiliation{Department of Physics and Astronomy, The University of Manchester, Manchester M13 9PL, UK}
\author{M. J. Doyle}
\affiliation{Department of Physics and Astronomy, The University of Manchester, Manchester M13 9PL, UK}
\author{W. Guo}
\affiliation{Mechanical Engineering Department, Florida State University, Tallahassee, FL 32310, USA}
\affiliation{National High Magnetic Field Laboratory, Tallahassee, FL 32310, USA}
\author{P. M. Walmsley}
\affiliation{Department of Physics and Astronomy, The University of Manchester, Manchester M13 9PL, UK}
\author{A. I. Golov}
\email{Author to whom correspondence should be addressed: andrei.golov@manchester.ac.uk}
\affiliation{Department of Physics and Astronomy, The University of Manchester, Manchester M13 9PL, UK}

\date{\today}

\begin{abstract}
We describe a versatile apparatus for optical observations of experimental processes at temperatures down to 0.1\,K. The cooling is achieved by a wet cryostat with a \3-\4 dilution refrigerator on a vibrationally-isolated platform, capable of continuous rotation at angular velocity of up to 3\,rad\,s$^{-1}$. The illumination light beam from lasers on a non-rotating optical table at room temperature is introduced via an optical fiber. The images are transferred to the intensified camera at room temperature through a coherent bundle of $10^5$ optical fibers limiting the spatial resolution to $\sim 30$\,$\mu$m, depending on the magnification used. The adjustment of the position of the illumination light, as well as of the focusing of the camera on the object under investigation, can be controlled remotely with the help of piezoelectric positioners. The apparatus was used for visualization of particles dispersed in superfluid helium at temperatures down to 0.14\,K. In one version of experiment, fluorescent light from clouds of excimer molecules He$_2^*$, generated in liquid helium by electron impact from electrons injected by field-emission tips, was recorded. In another, fluorescent particles of mean radius 3.0\,$\mu$m were initially loaded onto the horizontal surface of a piezoelectric crystal of LiNbO$_3$ and then injected into liquid helium by bursts of high-amplitude oscillations at the crystal’s resonant frequency 1\,MHz. The trajectories of particles in superfluid helium could be recorded at a frame rate of up to 990\,fps. The effective radius of particle images is $\sim 60$\,\um, while the uncertainty on the positions of their centers is $\sim 5$\,\um.  
\end{abstract}

\maketitle

\section{Introduction}
Dynamic tangles of quantized vortex lines in superfluid helium, also known as Quantum Turbulence, are an emerging counterpart of turbulence in classical fluids. While sharing certain common features, such as the Kolmogorov spectrum at larger lengthscales and applicability of the fruitful concept of inertial cascades of energy through lengthscales, they also possess interesting differences associated with the singular nature of the vorticity \cite{Vinen2002,QTBook2025}. 
The main ingredients of the dynamics of quantum turbulence are reconnections  and consequent helical deformations (Kelvin waves) of interacting quantized vortex lines. Theoretical modeling \cite{TsubotaNemirovskii2000} predicts that with decreasing temperature $T$ below some 0.5\,K, where the viscous damping effectively vanishes, the Kelvin waves become highly underdamped, and new regimes are expected.
These include the proliferation of Kelvin waves of all lengthscales including very short ones, the transfer (cascade) of energy towards shorter, dissipative lengthscales via interacting Kelvin waves \cite{Svistunov1995,Vinen2003}, frequent self-reconnections of vortex lines resulting in the emission of small vortex loops \cite{KS,Baggaley}, etc. 

In previous experiments, certain elementary processes involving discrete vortex lines in superfluid \4 at $T \gtrsim 1.3$\,K have been visualized using suspensions of small tracer particles: counterflowing superfluid and normal components \cite{VisCounterflowVanSciver,BewleyFirstHidrogenVis}, anomalous statistical distributions of the velocity of particles interacting with vortex lines \cite{VelStatLathrop,LaMantia2014,Mastracci2018,Svancara2021}, reconnections of pairs of vortex lines \cite{VisReconnectLathrop}, isolated vortex rings \cite{VisRingsGuo}, driven Kelvin waves on vortex lines \cite{VisReconnectAblation,VisKelvinGrenoble}.
To seek evidence for the novel behavior at lower temperatures, we built an apparatus in which micron-sized particles could be dispersed inside superfluid \4 at temperatures down to 0.1\,K, and their trajectories could be recorded using optical visualization. 
  Doing this at temperatures below 1\,K, which requires placing the experimental cell with superfluid helium into the vacuum can of a \3-\4 dilution refrigerator with much limited cooling power, poses several challenges: the impossibility of using the same means of introducing solid tracers into liquid helium as in experiments at higher temperatures (like the injection of gaseous helium-hydrogen mixture \cite{BewleyFirstHidrogenVis} or laser ablation of a solid target \cite{VisReconnectAblation}), limitation on the power of illuminating light and constraints on the types of optical access.

  \begin{figure}[h]
\includegraphics[width=10cm]
{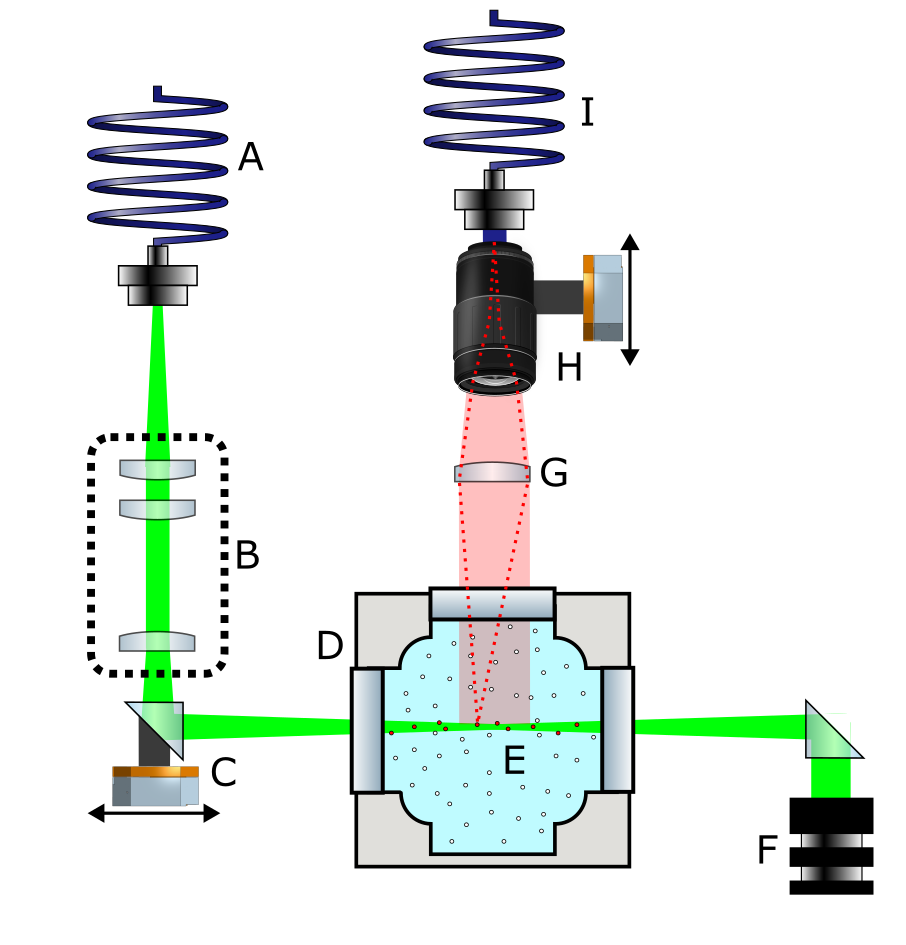}
\caption{
A schematic of the excitation and imaging optics and the experimental cell. This is the setup used for experiments with  excimers and early tests with microspheres: the illuminating light sheet has horizontal profile, and the imaging is through the top window of the cell. In later experiments with microspheres, the light sheet had vertical profile, and the imaging was through a side window. (A) multimode fiber, (B) combination of spherical and cylindrical lenses, (C) movable prism for shifting the light sheet, (D) experimental cell, (E) imaged region inside the cell, (F) dump, (G) achromat lens, (H) movable objective for focusing of the illuminated region onto the fiber bundle, (I) coherent fiber bundle.
}
\label{CellEtc}
\end{figure}

To overcome these challenges, we combined several innovations in a single apparatus (see Fig.\,\ref{CellEtc}): \newline
- the experimental cell, filled with superfluid $^4$He, with three optical windows was cooled by a $^3$He-$^4$He dilution refrigerator, capable of maintaining temperatures down to 0.1\,K;\newline
- the cryostat was placed on a vibrationally-isolated rotating platform while its gas-handling system was placed on another, co-rotating platform  \cite{Fear} -- thus allowing us to continuously rotate the experimental cell with superfluid helium and hence, generate and perturb arrays of parallel vortex lines of known density;\newline
- in situ dispersing fluorescent tracer particles of two different types: either He$_2^*$ excimer molecules \cite{excimers} or micron-size Fluoro-Max™ polymer microspheres \cite{Fluospheres}; \newline
- high-intensity pulses of light from lasers on a non-rotating laboratory frame were introduced into the vacuum can of the rotating cryostat using an optical fiber connected through a fiber-optic and electric rotary union \cite{RotaryUnion} and vacuum feedthrough; \newline
- inside the vacuum can, lenses formed the laser light into a light sheet, which illuminated the region of interest inside the experimental cell while passing through it and terminated in the thermally-isolated dump inside the vacuum can;\newline
- the image of the fluorescent particles was focused in situ onto a coherent bundle of $10^5$ optical fibers \cite{CoherentBundle} which transferred this image to the intensified camera outside the cryostat;\newline
- both the positioning of the light sheet inside the cell and focusing on it by the camera could be adjusted remotely using piezoelectric nanopositioners \cite{Attocube}.

\section{Experimental details}

\subsection{Experimental cell and cryogenics}

The helium cell, made of a ConFlat DN16CF 6-way cube with an inner volume of 13\,cm$^3$, was thermally linked to the mixing chamber (MC) of a customized rotating dilution refrigerator Oxford Kelvinox MX250 \cite{Fear}. 
The three antireflection-coated windows, with knife-edge rims, were sealed into DN16CF flanges using copper gaskets. To maximize the transmission of the illuminating light sheet, the corresponding pair of windows was made from either sapphire for the excimer He$_2^*$ excitation at 905\,nm or fused silica for the fluorescent particles excited at 532\,nm.

The base temperature of the MC before installing  the experimental cell and optical components was 12\,mK, while currently (whether with the cell empty or full of superfluid helium) it is 61\,mK. The cell is suspended from the MC on a sequential combination of two copper bars; their substantial total length of ($\sim 30$\,cm) was dictated by the need to accommodate the numerous optical components below the MC inside the limited space of the vacuum can. The dominant causes of the temperature difference between the cell and the MC were likely the thermal resistance of the several compressional connections in the thermal link and the radiation heat load from surrounding optical components at 4.2\,K. Currently, the base temperature of the full cell is 139\,mK, which is sufficient for experiments in the zero-temperature limit of quantum turbulence ($T \lesssim 300$\,mK), but it can likely be further decreased with future modifications of the thermal link.

\begin{figure}[h]
\includegraphics[width=12cm]
{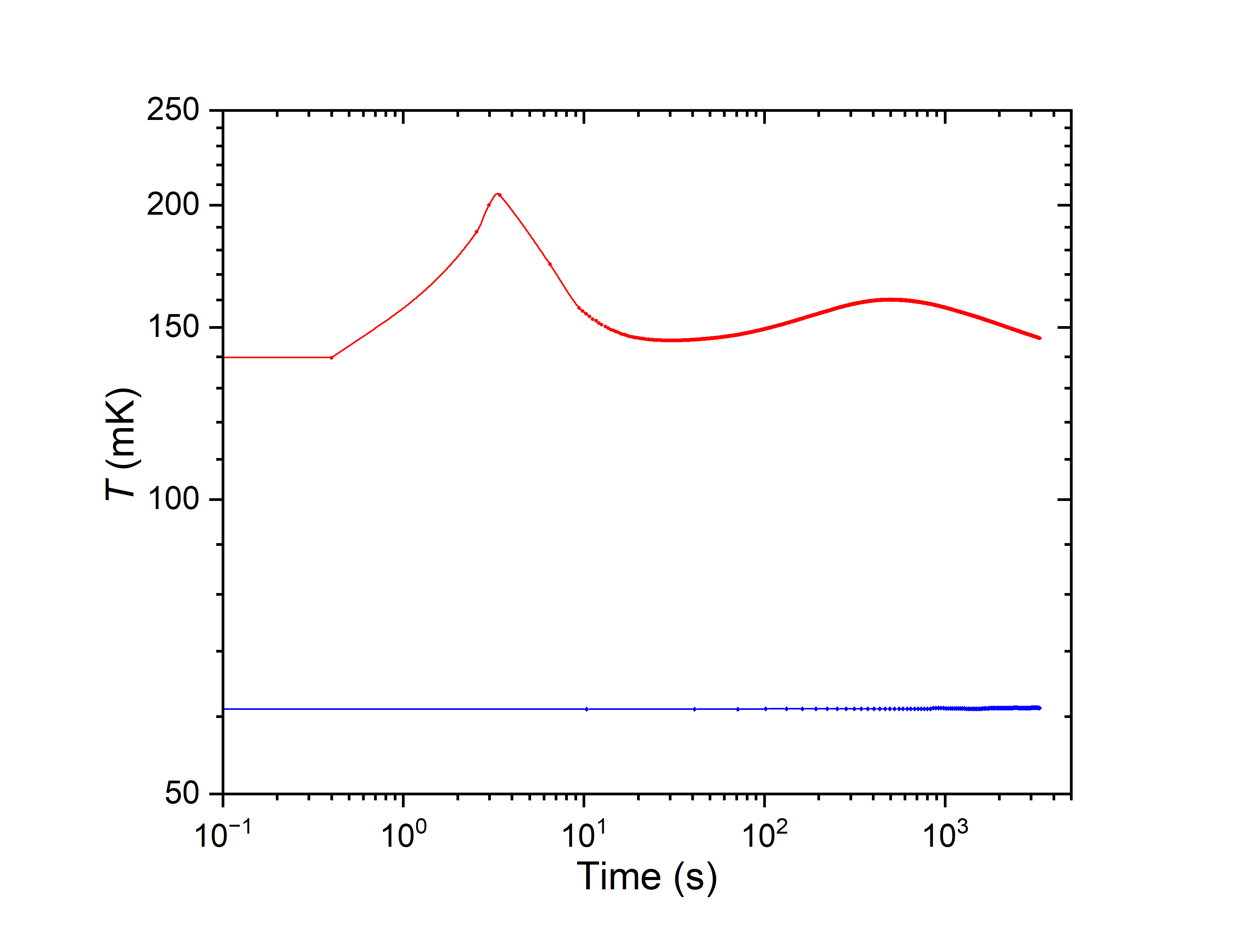}
\caption{Temperatures of the cell (top curve, red line and symbols) and of  the mixing chamber (bottom curve, blue line and symbols) after firing a 100\,$\mu$s pulse of 1.0\,kV voltage at the piezo transducer at low temperature.}
\label{TemperatureResponse}
\end{figure}

Temperatures were monitored using calibrated resistance thermometers read by two Lake Shore resistance bridges. One bridge was used for slow sequential monitoring of all stages of the refrigerator, while the other was in control of the cell at the highest read-out frequency. In Fig.\,\ref{TemperatureResponse}, an example is shown of the time dependence of the temperatures of the experimental cell and the MC upon applying a 0.1\,ms-long pulse of 1.0\,kV AC voltage to the piezo transducer for injecting particles into helium (see Section\,\ref{SectionParticles} for details). The temperature of the MC barely changes while the cell first responds with a temperature surge to slightly above 200\,mK at $\sim 3$\,s before relaxing towards the initial temperature after $\sim 10$\,s -- followed by some long-term oscillation of magnitude $\sim 10$\,mK on the timescale of an hour, likely caused by a redistribution of temperatures within the refrigerator.

\begin{figure}[h]
\includegraphics[width=12cm]
{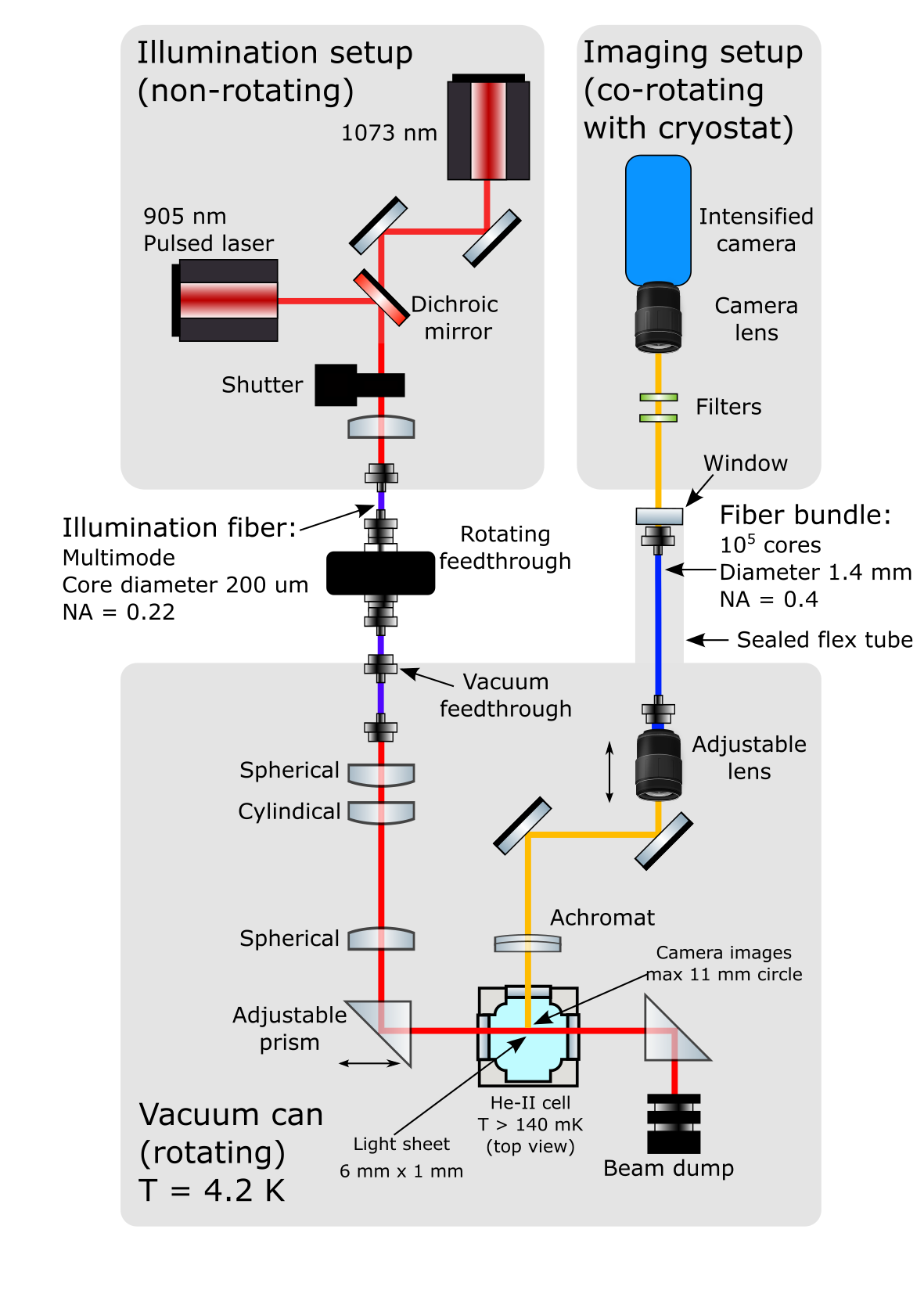}
\caption{Elements of the optical set up, grouped by their location by grey rectangles. The two groups at the top are at room temperature outside the cryostat; the left one on a stationary optical table, while the right one is mounted on top of the cryostat. The bottom group is inside the vacuum can of the dilution refrigerator.}
\label{SetUp}
\end{figure}

\subsection{Optical setup}
\label{section2p2}
All major elements of the optical set up are shown in Fig.\,\ref{SetUp}. The beams from lasers on an optical table were combined as necessary and coupled to a multimode optical fiber. The latter is fed into the rotating cryostat via a sequence of an optical rotary joint and vacuum feedthrough. Inside the cryostat, the beam exiting the fiber is shaped into an elliptical sheet of either $6{\rm \,mm} \times 0.8$\,mm (in experiments with He$_2^*$ excimers) or $11{\rm \,mm} \times 1$\,mm (with microspheres)
using an anamorphic system consisting of one cylindrical and two spherical antireflection coated N-BK7 lenses -- before entering the cell with superfluid $^4$He. After crossing the cell, the beam is diverted into the beam dump, thermally anchored to the helium bath of the cryostat at $4.2$\,K. The location of the light sheet inside the cell and focusing of the imaging lens could be adjusted remotely with the help of prisms mounted on piezo translation stages \cite{Attocube}. All optical components inside the vacuum can were anchored to the 4.2\,K stage. 

The pulsed laser unit was EKSPLA 252NT, allowing generation of $\sim 5$\,ns long pulses with wavelength tunable in the range 670--2600\,nm, the maximum pulse energy of 2\,mJ and pulse repetition rate of 1\,kHz. This provides a sufficient power density  (2.5\,mJ\,cm$^{-2}$ with a comparable pulse length \cite{excimers}) for He$_2^*$ fluorescence excited at 905\,nm. To speed-up the relaxation of excited vibration states of He$_2^*$ to the ground state, an auxiliary continuous 1073\,nm laser beam of intensity 3\,W\,cm$^{-2}$ was added. When using fluorescent microspheres as tracers, only pulsed light of wavelength 532\,nm was used for illumination.

The light from fluorescence emerging from the cell is focused, after passing through a sapphire window, on the coherent fiber bundle \cite{CoherentBundle} using a system of lenses. To minimize aberrations and maximize light collection by the fiber bundle with its numerical aperture NA whilst maintaining the desired spatial resolution, a four lens imaging system was designed. In a typical experiment we imaged an illuminated layer of liquid of between $\sim 6$\,mm and 
$\sim 10$\,mm in diameter onto the face of the fiber bundle of diameter 1.4\,mm. The required magnification $M$ is hence between approximately 0.25 and  0.14. 
The ambient temperature end of the fiber bundle is placed in close proximity to the sealed window from the vacuum can, facing the camera at the top of the cryostat. The emerging image is then focused onto the sensing element of the camera. 

With a magnification of $M=0.14$, the diffraction blur at the receiving end of the coherent fiber bundle is only $\sim M\times\frac{1.22\lambda}{D}L \approx 0.6$\,\um\ (here the wavelength is $\lambda = 640$\,nm, the achromat lens's diameter and distance from the illuminated region are $D=25$\,mm and the $L=130$\,mm, respectively). The actual resolution of the size of particles was dictated by the distance between the neighboring fibers of $4.2$\,\um, with further broadening due to occasional overlap of the particle's image on neighboring fibers as well as from the optical cross-talk between neighboring fibers. Taking the distance $4.2$\,\um\ as a conservative lower limit on the resolution at the receiving end of the fiber bundle, it would correspond to the actual distance inside the cell of 4.2\,\um$/M$, which is 30\,\um\ for $M=0.14$. At the camera end of the bundle of $10^5$ fiber cores when projected onto the camera's matrix area of diameter of about 800 pixels, each core is imaged by at least 2 pixels across, thus not further affecting the resolution much. We conclude that the size and shape of particles of diameter smaller than some 30\,\um\ could not be resolved; they should all appear as bright spots of diameter of at least 2 pixels, and only their cumulative brightness could be used as a proxy of the particle's size. Yet, computing the location of the brightness's centroid allowed to reduce the actual effective uncertainty of particle positions to $s_i\sim 5$\,\um\ (see Section\,4).  
It should be noted that, with particles moving at velocity $v_p$ exceeding $\sim s_i f \sim 0.5$\,cm\,s$^{-1}$ (estimated here for the frame rate $f = 1000$\,s$^{-1}$), the smallest resolved length scale of the physical property under study becomes limited by $s_f \sim v_p/f$ \cite{Svancara2021} and not by the particle image resolution $s_i$, quoted above.

The camera PCO dicam C1 had an intensiﬁer coupled to an sCMOS sensor with $2048 \times 2048$ pixels and a full frame rate of 200\,fps. The initial photocathode has quantum efficiency QE of 45\% at 640\,nm, so around half of all incoming photons were detected. The minimum gating time is 4\,ns which is well below the $\sim 100$\,ns radiative lifetime of the He$_2^*$ \cite{excimers}. The GaAsP photocathode has a diameter of 25\,mm requiring a magnification of circa 18 when imaging the fiber bundle. With 100 exposures of 200\,ns per frame we expected only around one dark electron per frame. Videos were recorded at 200, 415 or 990 fps with images of $1000\times1000$, $500\times1000$ and $200\times1000$ pixels, respectively. 

The pixel size was calibrated using a square grid with a 1\,mm period which was placed in the empty cell at room temperature, in the focal plane of the imaging system. The average line spacing in pixels was then measured, resulting in the ratio of 12.4\,\um/px (for the setup used in experiments with microspheres). An arrow on a secondary grid was used to establish the orientation of the images relative to the cell.

\section{Excimers}

The first particles investigated were He$_2^*$ excimers \cite{excimers} in the triplet state with a lifetime of 13\,s -- which can be excited by sequential absorption of two photons of wavelength 905\,nm followed by spontaneous emission (fluorescence) at 640\,nm. Thanks to their considerable effective radius of $0.6$\,nm one would expect them to stay trapped on vortex lines at temperatures below $\sim 0.6$\,K \cite{ZmeevTrapped} with the trapping diameter of $\sim 100$\,nm, although the trapped lifetime could be severely shortened by their collisions with other trapped He$_2^*$ molecules and $^3$He impurities. He$_2^*$ can be created by an electron impact.  
Our main interest in using them as vortex tags was in that they are atomically small (i.\,e. the least invasive), can be created in situ and then disappear without leaving contamination, and are also electrically neutral (unlike positive and negative ions of comparable size, which are frequently used to manipulate and detect the presence of vortex lines \cite{NegativePositiveIons,NegativeIonsTangle}).

The excimers were produced by electron impact from electrons injected into helium from a sharp tungsten tip \cite{GolovIshimoto} at voltage $\sim 500$\,V. There were two tips mounted on opposing flanges at right angles to both the excitation and fluorescence beam. Each could be used to generate a vortex tangle \cite{NegativeIonsTangle} and produce excimers \cite{ZmeevTrapped}. Each tip had two 97.3\% transparency tungsten meshes mounted in front of them separated by 1\,mm. The closest grid to the tip was an independent electrode whilst the farther grid was grounded to the cell body. 

\begin{figure}[h]
\includegraphics[width=10cm]{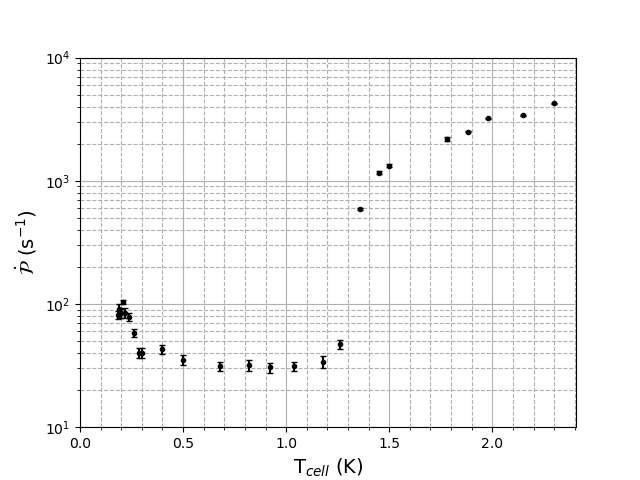}
\caption{Received photon flux vs. temperature for excimers generated by pulses of electron currents of order 1\,nA for 0.2\,s.}
\label{T-dependence_excimers}
\end{figure}

Tracing individual excimers requires receiving and recording, within the limited NA of our telescope, quantum efficiency of the camera and background noise, of photons from the same molecule in each consecutive video frame (which might be possible in principle but only for long exposures with some 1000 excitation-emission cycles per frame). Instead, in our experiments, we measured the total number of photons from excimers, recorded by the camera from the given solid angle and imaged area in the given time, which is shown versus temperature in Fig.\,\ref{T-dependence_excimers}.  

At the temperatures at which excimers are expected to be permanently bound to vortex lines, $T\lesssim 0.6$\,K, they move ballistically through helium, and the cross-section for trapping on vortex lines is small. Furthermore, even if trapped, their lifetime becomes greatly shortened by the presence of other trapped excimers and $^3$He atoms, which concentrate at the cores of quantum vortices at low temperatures. 
We did use very pure $^4$He from Peter McClintock's laboratory in Lancaster with the apparent $^3$He concentration of somewhere between $2\times 10^{-10}$ -- $2\times 10^{-15}$ (not precisely known \cite{Hendry1987,HendryLT17,McClintock2022}). Unfortunately, no detectable densities of excimers, trapped on vortex lines, have been observed in our experiments at any temperatures down to 0.2\,K. 

\section{Solid particles}
\label{SectionParticles}

Having little success with excimers as vortex markers, we turned to solid particles (microspheres) \cite{Fluospheres} dispersed in situ. These were dyed polysterene spheres Fluoro-Max of density 1.06\,g\,cm$^{-3}$ and mean radius $a=3.0$\,$\mu$m (with uniformity $<18$\%, implying that most particle radii are within $\pm 0.5$\,\um) which can be excited by the light of a convenient wavelength 532\,nm and then emit photons at circa 640\,nm. As they are not neutrally buoyant, the particles are constantly under the vertical force due to gravity, 
yet they can still yield information about the structure and dynamics of the turbulence in both normal and superfluid components. Thanks to the substantial binding energy, they are expected to be trapped by vortex lines, although their large weight and strong interaction with thermal excitations (`the normal component') of superfluid helium might cause intermittent liberation \cite{Sergeev}. 

To disperse particles, a disk (12.7\,mm diameter, 1\,mm thickness) of a piezoelectic crystal $36^{\circ}$ Y-cut LiNbO$_3$ \cite{BostonPiezoOptics} with resonance at 1.014\,MHz of $Q=500$ (when in liquid $^4$He) at the bottom of the cell was initially covered with many layers of dry particles. These could be jettisoned  into liquid helium by applying a short (100--300\,$\mu$s) burst of AC voltage of the resonant frequency and amplitude 0.8--1.5\,kV. This method of dispersing particles in superfluid helium \cite{JinMaris} was pioneered in the group of Humphrey Maris following their extensive research on the use of piezo transducers in liquid helium. Unlike the method \cite{Lathrop} utilizing vibrations from a powerful room-temperature ultrasonic vibrator, suitable for helium-bath cryostats at temperatures $\sim 2$\,K, high-Q piezo transducers allow application of large-amplitude vibrations inside a $^3$He-$^4$He dilution refrigerator at temperatures down to 0.1\,K and lower.

Compared to excimers, these particles are much brighter even though they require much lower light intensities for the excitation of fluorescence. For illumination, we used  laser light of wavelength 532\,nm and power $\sim 100$\,mW, since with higher intensities a temperature gradient in the cell would induce a horizontal current of the normal component (counterflow), which in turn caused a horizontal drift of particles towards the exit window. This temperature gradient was likely caused by stray illumination light arriving at the rim of  the entry window. Future improvements will incorporate a mask, blocking off all stray light around the central rectangular region of the otherwise elliptical laser sheet. This will also assure a more uniform illumination of particles across all the imaged region. Additionally, an immersion heater could be placed around the outer window to allow the temperature gradient to be varied in a controlled way (including compensating for the nuisance temperature gradient if necessary).

We were able to disperse particles at all temperatures between 0.2\,K and 1.2\,K.  
During about 1\,s following the injection large-scale turbulent eddies were evident. At later times they decayed, and simple settling of particles could be monitored. However, in both regimes, certain  particles did not follow similar smooth trajectories of the majority of their neighbors but moved either with different velocity and direction or had quite erratic trajectories. We associate this behavior with the particles being trapped on vortex lines. Also, there were cases of very bright objects, which we associate with clumps of several  particles -- these were usually moving slowly and often erratically, most likely being trapped by one or more vortex lines. 

\begin{figure}[h]
\includegraphics[width=15cm]{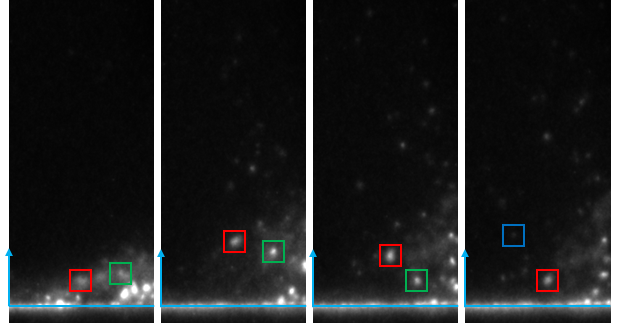}
\caption{Images of particles at $T=1.00$\,K at times (left to right) 1\,ms, 10\,ms, 20\,ms and 27\,ms after the beginning of the injection burst of duration 0.1\,ms and AC voltage of 1.0\,kV. The horizontal light blue line indicates the surface of the piezo transducer loaded with particles at elevation $y=0$. The vertical light blue arrow has length of 1\,cm. The green and red squares mark two clumps of particles, whose trajectories $y(t)$ are analyzed in Section 4. The blue square marks an individual particle.}
\label{FourFrames}
\end{figure}

\begin{figure}[h]
\includegraphics[width=15cm]{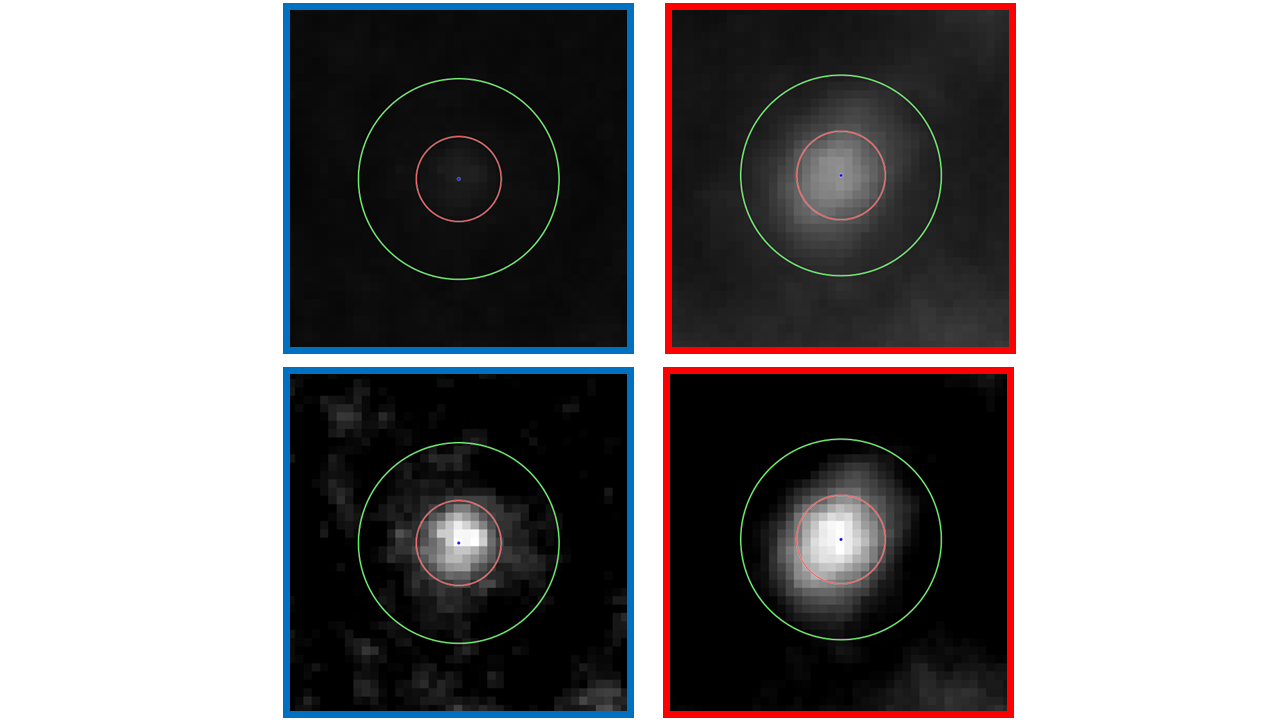}
\caption{Zoomed images, each of a 40\,px = 0.5\,mm side, of two objects from the right panel of Fig.\,\ref{FourFrames}. Left -- a single particle, right -- a clump of particles. Top row -- raw images, bottom row -- enhanced images (after subtracting the local background and normalizing to the brightest pixel). The blue dot in the middle of each image is the location of the centroid of the brightness of pixels bound by the green circle of radius $w=12$\,px. The effective radius  of each image is shown by pink circles: $m_2^{1/2} =$ 5.1\,px = 63\,\um\ for the left and $m_2^{1/2} = 5.3$\,px = 66\,\um\ for the right one. The objects' luminosities $m_0$ (with the dynamic range of each pixel normalized to unity) are 2.12 and 11.7 for the left and right images, respectively.}
\label{FourCrops}
\end{figure}

In Fig.\,\ref{FourFrames}, examples are shown of four instances of an evolution of particle's positions after the injection. 
There are many different ways how the obtained images can be analyzed (examples are the conventional PIV and PTV techniques of measuring the particle velocities and accelerations \cite{Mastracci2018,Svancara2021}). We are currently in the process of developing some of these analytical techniques in connection with particle tracking algorithms, stable in the presence of  background noise, weak contrast and broad distributions of particle luminosities and velocities.

\begin{figure}[h]
\includegraphics[width=20cm]{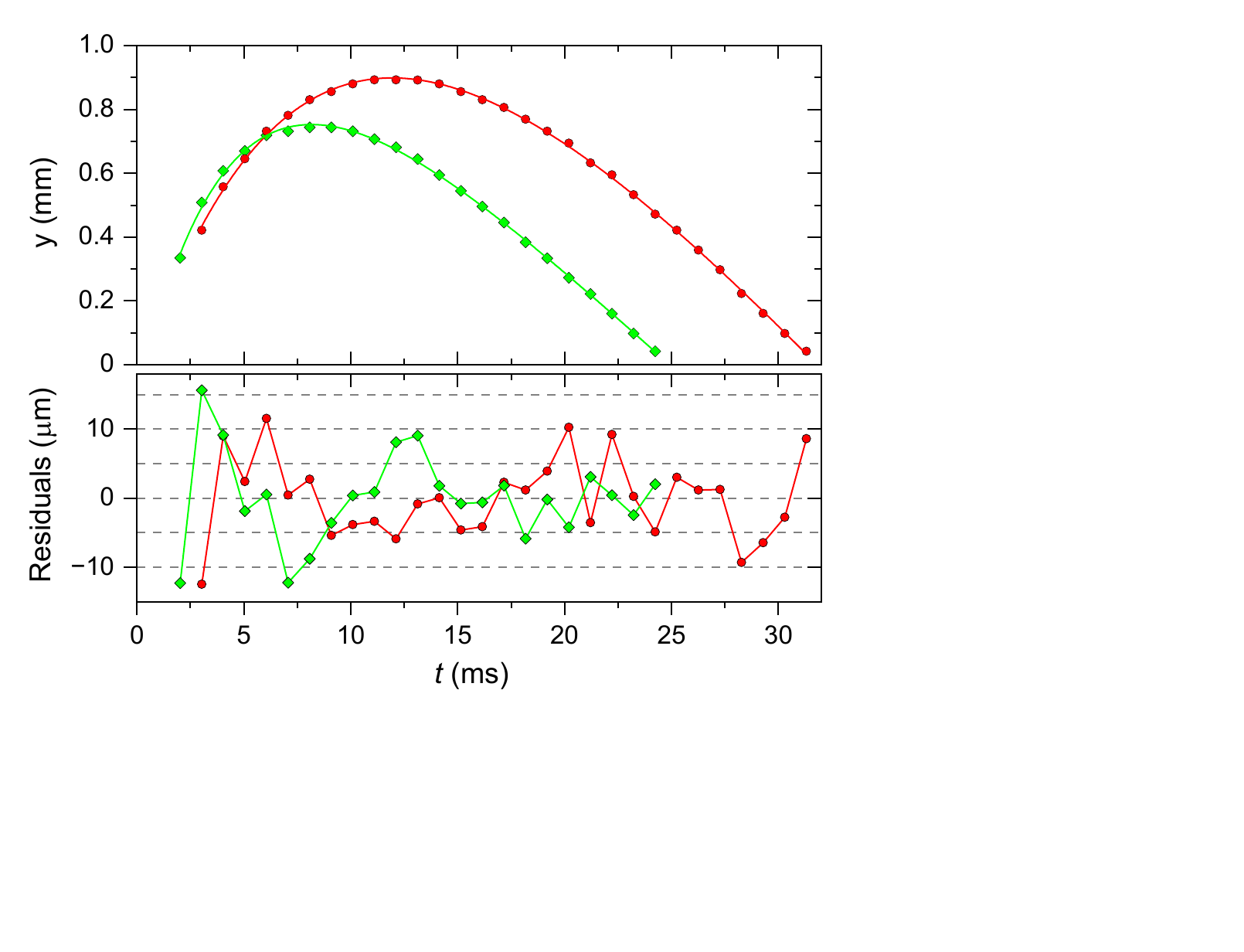}
\caption{Top panel: vertical components of trajectories of two clumps of particles $y(t)$ at $T=1.00$\,K taken at 990\,fps, fitted by Eq.\,\ref{ExpVelRelaxY}. Bottom panel: residuals of the fitting; same symbols for the two corresponding datasets, solid lines connect the data points.}
\label{SmoothResidualsY}
\end{figure}

To characterize the images of individual objects in terms of their position $(x,y)$, luminosity $m_0$ and effective radius $m_2^{1/2}$, we followed the method of Crocker and Grier \cite{CrockerGrier1996}. Namely, the zeroth, first and second moments of the brightness of pixels $A(i,j)$ were computed within a circle of radius of $w=12$\,px around a guessed center of the object $(x_0,y_0)$: 
\begin{equation}
m_0 = \sum_{i^2+j^2 \le w^2} A(x_0+i,y_0+i),
\label{m0}
\end{equation}
\begin{equation}
(\delta x,\delta y) = \frac{1}{m_0}\sum_{i^2+j^2 \le w^2} (i,j)A(x_0+i,y_0+i),
\label{m1}
\end{equation}
\begin{equation}
m_2 = \frac{1}{m_0}\sum_{i^2+j^2 \le w^2} (i^2+j^2)A(x_0+i,y_0+i).
\label{m2}
\end{equation}
If the computed displacement of the centroid $(\delta x,\delta y)$ were greater than half a pixel, the procedure repeated around an updated center $(x_0,y_0) := (x_0+\delta x,y_0+\delta y)$ until converging. This resulted in a sub-pixel precision of the object's location. In Fig.\,\ref{FourCrops}, we compare images of two objects: a single particle and a clump of particles. While their luminosities ($m_0=2.12$ and $m_0=11.7$, respectively) differ by a factor of 5.3, the effective radii ($m_2^{1/2}=5.08$\,px = 63\,\um\ and $m_2^{1/2}=5.30$\,px = 66\,\um) are nearly identical, about twice the conservative estimate on the spatial resolution of $\sim 30$\,\um\ from Section \ref{section2p2}  -- confirming that it is limited by the distance between  individual fibers in the optical bundle. 

  To demonstrate the accuracy of tracing particle positions, we chose two bright clumps of particles with smooth trajectories after they were launched upwards at temperature $T=1.00$\,K at time $t=0$ and imaged at 990\,fps (these are the objects marked by red and green squares in Fig.\,\ref{FourFrames}). The time dependence of their vertical coordinates, $y(t)$, is shown in the top panel of Fig.\,\ref{SmoothResidualsY}. Solid lines represent the fitting by the formula 
  \begin{equation}
  y=v_s(t-t_0)+\tau(v_0-v_s)\left(1-\exp(-(t-t_0)/\tau)\right),
  \label{ExpVelRelaxY}
  \end{equation}
  with the values of the fitting parameters $t_0$ (effective time delay after applying the AC voltage burst to the piezoelectric transducer), $v_0$ (initial vertical velocity), $v_s$ (terminal slip velocity) and $\tau$ (Stokes relaxation time) listed in Table\,\ref{tab1}. The quality of fitting by such a simple linear model as Eq.\,\ref{ExpVelRelaxY} is impressive (as we mentioned above, most particles move on less regular trajectories). The value of the effective initial time offset $t_0 \sim 0.3$\,ms is comparable with the duration of the injection burst (the piezo crystal with $Q \sim 500$ only reaches its largest amplitude of oscillations at the end of the 0.1\,ms-long burst which contains 100 cycles at frequency of 1\,MHz); it also absorbs the uncertainty of the actual initial elevation of the clump within many layers of particles on the surface of the transducer, $y(0) \sim v_0t_0 \lesssim 100$\,\um, which was rigidly set to $y(0)=0$ in Eq.\,\ref{ExpVelRelaxY}.
    
  The residuals are shown in the bottom panel of Fig.\,\ref{SmoothResidualsY}: there is little low-frequency structure left while the random component of residuals has amplitude of order 5\,\um. Since we do not know the extent of the contribution from the hydrodynamic fluctuations into these deviations form a smooth trajectory, we take $s_i \sim 5$\,\um\ for the upper estimate on the accuracy of the spatial resolution of particle coordinates. This value is comparable with half of the pixel size (6.2\,\um), and is about an order of magnitude smaller than the effective radius of particle images $\sim 60$\,\um, limited by the size of individual fibers in the optical bundle.

  \begin{table}[h]
\begin{center}
\begin{tabular}{|c|c|c|c|c|c|c|c|}
\hline Symbol & $t_0$(ms) & $v_0$(cm\,s$^{-1}$) & $v_s$(cm\,s$^{-1}$) & $\tau$(ms) & $R_*$(\um) & $\tau_*$(ms) & ${\mathcal Re}_n$
\\
\hline red $\circ$ & 0.27 & 19.4 & -7.7 & 9.3 & 12.2 & 5.9 & 1.24
\\
  green $\diamond$ & 0.39 & 26.3 & -6.2 & 4.7 & 11.0 & 5.1 & 1.52
 \\
\hline
\end{tabular}
\end{center}
\caption{Fitting parameters $t_0$, $v_0$, $v_s$ and $\tau$ for the datasets from Fig.\,\ref{SmoothResidualsY} fitted by Eq.\,\ref{ExpVelRelaxY} and parameters $R_*$, $\tau_*$ and $\mathcal Re$ inferred using Eqs.\,\ref{vs}, \ref{tau_*} and \ref{Re_n}.}
\label{tab1}
\end{table}

Let us assume that the clump is nearly spherical with radius $R_*$ and density $\rho_p$, and the dominant contribution into the drag force comes from the Stokes drag. Then 
  the values of the terminal velocity, Stokes relaxation time and highest Reynolds number over the trajectory should meet:
   \begin{equation}
  |v_s| = \frac{2(\rho_p-\rho)g}{9\eta_n}R^2,
  \label{vs}
  \end{equation}
   \begin{equation}
 \tau = \frac{\rho_p-\rho}{\rho_p+\rho/2}\frac{|v_s|}{g} ,
 \label{tau_*}
  \end{equation}
    \begin{equation}
{\mathcal Re}_n = \frac{\rho_n v_0 (2R_*)}{\eta_n},
 \label{Re_n}
  \end{equation}
 where $\rho = 145$\,kg\,m$^{-3}$ is the density of liquid helium, $\rho_n = 1.02$\,kg\,m$^{-3}$ and $\eta_n = 3.87\times 10^{-6}$\,Pa\,s are the density and viscosity of the normal component at $T=1.00$\,K \cite{Carlo}, and $g=9.81$\,m\,s$^{-2}$ is the gravitational strength. 

 The values of the effective radii of a clump $R_*(v_s)$ from Eq.\,\ref{vs}, the relaxation time $\tau_*(v_s)$ from Eq.\,\ref{tau_*} and the highest Reynolds number ${\mathcal Re}_n(v_0,R_*)$ are given in Table\,\ref{tab1}. The radius suggests that bright clumps contain of order $(R_*/a)^3 \sim 50$--70 individual particles, which seems a plausible upper estimate since real clumps are likely neither spherical nor densely packed. The lower estimate on this number comes from the ratio of luminosities $\sim 5.3$ of the larger clump and individual particle: with the assumption that the luminosities are proportional to the surface area (i.\,e. radius squared) of effectively-spherical objects, we arrive at $\sim 5.3^{3/2} = 12$ individual particles in the larger clump. The relaxation time, computed from $|v_s|$, is also in a reasonable agreement with the observed $\tau$. Finally, the Reynolds number of order unity validates the use of Stokes formula, Eq.\,\ref{vs} for the drag from laminar flow of hydrodynamic normal component around the particle. 
 Thus, our observations are consistent with the assumption that the viscous drag is essential at $T=1.0$\,K. Whether additional contributions to the drag force, such as that from quantized vortices, are important will require further investigation. 
 
\section{Conclusions}

We constructed a versatile apparatus with optical access to superfluid helium at temperatures down to $\sim 0.1$\,K, also in rotation. It is capable of taking photos of the light emitted by  fluorescent particles dispersed in superfluid \4. Images can be taken at up to 990 frames per second with the spatial resolution of particle positions of $\sim 5$\,$\mu$m. 

With clouds of excimer molecules He$_2^*$, only the mean local intensity of the emitted fluorescent light and not the positions of individual molecules could be recorded. The analysis of the absorption spectroscopy of He$_2^*$ at temperatures down to 0.2\,K and of the dynamics of the propagation of a jet of the normal component, decorated by He$_2^*$, at temperatures near 1\,K  will be published elsewhere. 

With micron-sized polymer particles, images of individual particles and their clumps at frame rate up to 990\,fps could be recorded. An analysis of smooth trajectories of clumps of several dozen particles, launched upwards and then relaxing towards settling at a constant terminal velocity at temperature $T=1.00$\,K, is mainly consistent with the assumptions that these clumps are not trapped by vortex lines and experience a linear Stokes drag from the viscosity of stationary normal component. The analysis of the trajectories of all types of particles, either trapped by vortex lines or free, at temperatures 0.2\,K--1.2\,K is ongoing. 

\begin{acknowledgments}
This work was largely inspired by our collaboration with Joe Vinen, who was eager to understand the dynamics of tangles of vortex lines in superfluid helium in the zero-temperature limit. 
The financial support came from EPSRC through grant No. EP/P025625/1. W. Guo acknowledges support from the Gordon and Betty Moore Foundation through grant DOI 10.37807/gbmf11567 and the National High Magnetic Field Laboratory at Florida State University, which is supported by the National Science Foundation Cooperative Agreement No. DMR-2128556 and the State of Florida. 
\end{acknowledgments}


\section*{Data Availability}

The data that support the findings of this study are available within the article. 


\end{document}